\documentclass[twocolumn,pre,showpacs,superscriptaddress]{revtex4-1}
\usepackage[left=2.5cm,right=2.5cm,top=2cm,bottom=3cm,includeheadfoot]{geometry}
\usepackage{graphicx}
\usepackage{rotating}
\usepackage{amsmath}
\usepackage{amsfonts}
\usepackage{amssymb}
\usepackage{enumerate}
\usepackage{longtable}
\usepackage{units}
\usepackage{hyperref} 
\usepackage[T1]{fontenc}
\usepackage[latin1]{inputenc}
\usepackage{graphicx}
\usepackage{capt-of}
\usepackage{bbm}
\usepackage{dcolumn}
\usepackage{bm}
\usepackage{color}

\begin{document}
\newcommand{\cred}{\color{red}}
\title{Extracting critical exponents for sequences of numerical data via series extrapolation techniques}

 \author{Kris C\"oster}
 \affiliation{Lehrstuhl f\"ur Theoretische Physik I, Otto-Hahn-Str.~4, TU Dortmund, D-44221 Dortmund, Germany}
 \email{kris.coester@tu-dortmund.de}
 
 \author{Kai Phillip Schmidt}
 \affiliation{Lehrstuhl f\"ur Theoretische Physik I, Staudtstra\ss e 7, FAU Erlangen-N\"urnberg, D-91058 Erlangen, Germany}
\email{kai.phillip.schmidt@fau.de}

\date{\rm\today}

\begin{abstract}
We describe a generic scheme to extract critical exponents of quantum lattice models from sequences of numerical data which is for example relevant for non-perturbative linked-cluster expansions (NLCEs) or non-pertubative variants of continuous unitary transformations (CUTs). The fundamental idea behind our approach is a reformulation of the numerical data sequences as a series expansion in a pseudo parameter. This allows to utilize standard series expansion extrapolation techniques to extract critical properties like critical points and critical exponents. The approach is illustrated for the deconfinement transition of the antiferromagmetic spin 1/2 Heisenberg chain. 
\end{abstract}


\maketitle

\section{Introduction}
\label{Sect:Intro}
Our understanding of correlated quantum-many body systems is an important topic of 
current research in physics, since fascinating types of complex quantum phases and 
associated collective behaviour are already known and more is to discover in the future. 
These systems are notoriously complicated to treat theoretically, which
 is especially true close to quantum critical points where quantum fluctuations become
  long-range due to the diverging correlation length. Often, theoretical approaches treat finite systems numerically and one relies on a proper scaling in system size towards the thermodynamic limit, e.g.~this applies to exact diagonalizations (ED), quantum Monte Carlo simulations or density matrix renormalization group (DMRG) calculations.  

This is different for other real-space approaches which work directly in the thermodynamic limit, but introduce other truncation parameters, e.g.~the perturbative order in high-order linked-cluster expansions (LCEs), the number of sites in NLCEs, or the spatial extension of operators in CUTs. Close to criticality, one aims at a extrapolation of all these parameters to infinity in order to cover the diverging real-space correlations properly. 

In many cases a simple scaling towards this limit is not known and also not expected. In contrast, powerful extrapolation schemes, corresponding to an extrapolation to infinite order in the perturbation parameter, exist in the established field of LCEs. Here, the linked cluster theorem is used to determine physical quantities in the thermodynamic limit by performing calculations on finite linked clusters. These extrapolation schemes represent sophisticated resummation techniques allowing to extract critical properties like critical points and, most importantly, critical exponents. 

The non-perturbative counterpart of LCEs is commonly referred to as numerical linked-cluster expansions or non-perturbative linked-cluster expansions (NLCEs). The essential idea behind all NLCEs is a non-perturbative treatment of graphs, achieved via an exact (block) diagonalization, yielding results in the thermodynamic limit after an appropriate embedding procedure. The underlying idea can be traced back to Irving and Hamer in 1984 who replaced the series expansion of the ground-state energy per cluster by the corresponding numerically exact value for this cluster using exact diagonalization \cite{Irving1984}. In principle, it is possible to modify all high-order series expansions via LCEs in this fashion. However, the power of this concept remained unheeded for some time but received more attention recently and many exciting developments have been achieved in this direction \cite{Rigol2006,Rigol2007_1,Rigol2007_2,Yang2011,Yang2012,Khatami2011,Kallin2013,Tang2013,Kallin2014,Stoudenmire2014,Ixert2014,Coester2015,Devakul2015}. Apart from different physical questions tackled by NLCEs, also different techniques to treat the systems {\it on} the graphs have been used. While most works applied ED (except Refs.~\cite{Kallin2014,Stoudenmire2014} using DMRG) on graphs, it has been established that non-perturbative CUTs on graphs (so-called gCUTs \cite{Yang2011}) give exciting new perspectives on NLCEs due to the additional freedom during the flow on the graphs \cite{Coester2015}.  

Overall, NLCEs represent a highly versatile approach. The reason for this lies in the simple requirements for this technique; the quantities of interest in the thermodynamic limit must exist on finite clusters. Consequently, NLCEs will probably also be applicable for many other challenges in the future. Nevertheless, powerful extrapolation schemes like in LCEs allowing to extract critical properties like critical points and critical exponents do not exist. In contrast to LCEs, NLCEs and all non-perturbative real-space approaches mentioned above provide sequences of numerical data points. While first extrapolation schemes are applied for NLCEs \cite{Tang2013, Kallin2013}, to the best of our knowledge, no similar extrapolation schemes for the extraction of critical exponents are currently available.

Here we propose a novel scheme to extrapolate such numerical data sequences. To this end the numerical data points are mapped to a series expansion in a pseudo perturbation parameter. As a result, one can implement the standard series expansion techniques, which gives, as we argue, access to critical points as well as critical exponents.

The paper is organized as follows. We describe our extrapolation scheme in Sect.~\ref{Sect:approach} and we apply it to the deconfinement transition of the spin 1/2 Heisenberg chain in Sect.~\ref{Sect:applications}. Finally, we give conclusions in Sect.~\ref{Sect:Conclusions}. 

%
%
\section{Extrapolation scheme}
\label{Sect:approach}
This section contains all general and technical aspects of our novel extrapolation scheme for numerical data sequences. We start by giving a brief introduction into the well-established extrapolation techniques for LCEs. For a more detailed overview over this vast topic, we refer the interested reader to the well-written introduction by Guttmann \cite{Guttmann}. On this basis, we describe how to use these techniques to extrapolate numerical data sequences from non-perturbative approaches like NLCEs.
\subsection{Extrapolating series expansions}
We consider a series expansion of the form
\begin{align}
F(\lambda)=\sum_{n\geq 0}^m a_n \lambda^n=a_0+a_1\lambda+a_2\lambda^2+\dots a_m\lambda^m,
\end{align}
with $\lambda\in \mathbb{R}$ and $a_i \in \mathbb{R}$. The function $F(\lambda)$ represents an approximant of the actual function \mbox{$\tilde{F}(\lambda)=\lim_{n\rightarrow \infty}F(\lambda)$}. Here, $\tilde{F}(\lambda)$ may represent the excitation gap, the entanglement entropy or any other quantity accessible via LCEs. Naturally, depending on the quantity and the value of $\lambda$, the approximation can become deficient.

The fundamental idea behind extrapolation schemes is the derivation of extrapolants from $F(\lambda)$. These extrapolants are functions whose form differ from the plain series expansion, which leads to a better approximation, i.e.~, the form of the extrapolant is generically more suited to mimic the behavior of the actual physical function $\tilde{F}(\lambda)$ than a plain series.

A standard extrapolation scheme is the Pad\'e extrapolation defined by 
\begin{align}
P[L/M]_{F}:=\frac{P_L(\lambda)}{Q_M(\lambda)}=\frac{p_0+p_1\lambda+\dots + p_L \lambda^L}{q_0+q_1\lambda+\dots q_M \lambda^M}\quad,
\end{align}
with $p_i\in \mathbb{R}$ and $q_i \in \mathbbm{R}$ and $q_0=1$. The latter can be achieved by reducing the fraction. The real coefficients are fully defined by the condition that the Taylor expansion of $P[L/M]_{F}$ about $\lambda=0$ up to order $L+M$ with $L+M\leq m$ recovers the corresponding Taylor expansion of the original series $F(\lambda)$.

Naturally, Pad\'e extrapolants are more versatile than a plain series and are specifically suited for scenarios where a rational function is approximated. Poles of an extrapolant can either reflect physics of the system or they can simply be an artifact of the extrapolation technique. If such a spurious pole is located close to or between $\lambda=0$ and the considered $\lambda$ value, the corresponding extrapolant is called \textit{defective} and should be discarded.\\
There is no single blueprint distinguishing physical and defective poles and this must be decided in the respective context and matched with the expectations. The extrapolation is considered to work if several combinations of $L$ and $M$ yield similar results. Specifically relevant is the convergence of the families defined by $L-M=\text{const}$. Pad\`e extrapolants constitute a valid extrapolation scheme, however, especially close to quantum criticality, the rational functions fail to capture the characteristic behavior.

In these cases, it is advisable to implement the so-called Dlog-Pad\'e extrapolation, which are applicable to quantities of definite sign like energy gaps or spectral weights. Most importantly, this scheme allows the extraction of critical exponents, i.e., the extrapolants are suited to describe power-law behavior.

If one assumes power-law behavior near a critical value $\lambda_{\rm c}$, the function $\tilde{F}(\lambda)$ close to $\lambda_{\rm c}$ is given by
\begin{align}
\tilde{F}(\lambda)\approx \left(1-\frac{\lambda}{\lambda_{\rm c}}\right)^{-\alpha} A(\lambda).
\end{align}
If $A(\lambda)$ is analytic at $\lambda=\lambda_{\rm c}$, we can write
\begin{align}
\tilde{F}(\lambda)\approx \left(1-\frac{\lambda}{\lambda_{\rm c}}\right)^{-\alpha}A|_{\lambda=\lambda_{\rm c}}\left(1+\mathcal{O}(1-\tfrac{\lambda}{\lambda_{\rm c}})\right).
\end{align}
Near the critical value $\lambda_{\rm c}$, the logarithmic derivative is then given by
\begin{align}
\tilde{D}(\lambda)&:=\frac{\text{d}}{\text{d}\lambda}\ln{\tilde{F}(\lambda)}\label{dx}\\
&\approx \frac{\alpha}{\lambda_{\rm c}-\lambda}\left\{ 1+ \mathcal{O}(\lambda-\lambda_{\rm c})\right\}\nonumber.
\end{align}
In the case of power-law behavior, the logarithmic derivative $\tilde{D}(\lambda)$ is expected to exhibit a single pole. In this case, Pad\'e extrapolations are perfectly suited to approximate $\tilde{D}(\lambda)$. These extrapolants are defined by the corresponding series $D(\lambda)$, which is, due to the derivative, only known up to order $m-1$, i.e., $L+M\leq m-1$. The resulting Dlog-Pad\'e approximants of $F(\lambda)$ are then defined by
\begin{align}
dP[L/M]_F(\lambda)=\exp\left(\int_{0}^\lambda P[L/M]_{D}\quad\text{d}\lambda'\right)
\end{align}
and represent physically grounded extrapolants in the case of a second-order phase transition. The poles of $P[L/M]_{D}(\lambda)$ can either indicate a critical value $\lambda_{\rm c}$ or be spurious. In practice, this is decided essentially by the location of these poles. The corresponding critical exponent of a pole $\lambda_{\rm c}$ is given by
\begin{align}
\alpha\equiv\left.\frac{P_L(\lambda)}{\tfrac{\text{d}}{\text{d}\lambda} Q_M(\lambda)}\right|_{\lambda=\lambda_{\rm c}} \label{extract_exponent}.
\end{align}
If the exact value of $\lambda_{\rm c}$ is known, one can obtain better estimates of the critical exponent by defining
\begin{align*}
\alpha^*(\lambda)&\equiv(\lambda_{\rm c}-\lambda)D(\lambda)\\
&\approx \alpha+\mathcal{O}(\lambda-\lambda_{\rm c}),
\end{align*}
where $D(\lambda)$ is given by Eq.(\ref{dx}). Then
\begin{align}
P[L/M]_{\alpha^*}\big|_{\lambda=\lambda_{\rm c}}=\alpha \label{biasnue}
\end{align}
yields a (biased) estimate of the critical exponent.

It should be noted that Dlog-Pad\'e extrapolants can also prove to be proper extrapolants if \textit{no} phase-transition is described. Finally, let us mention that there exist also other powerful extrapolation schemes for high-order series expansions like integral approximants which we do not detail here. 

\subsection{Extrapolating non-perturbative linked-cluster expansions}
\begin{figure}[t]
\begin{center}
\includegraphics*[width=0.85\columnwidth]{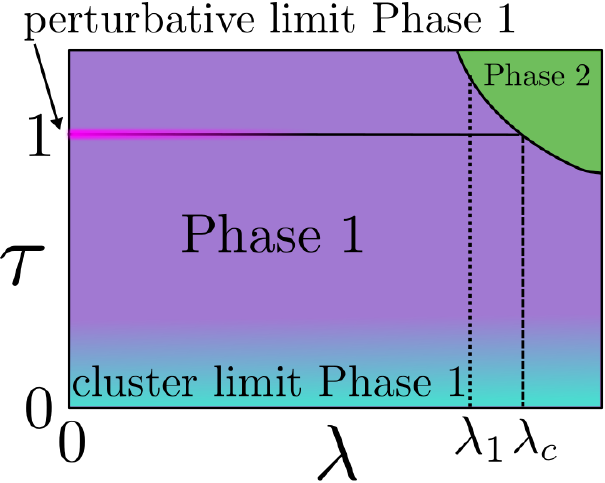}
\end{center}
\caption{Sketch of the anticipated phase diagram in the parameters $\lambda$ and $\tau$. The perturbative series expansion is carried out about the $\lambda=0$ limit on the physical axis ($\tau=1$). The NLCE expansion in $\tau$ is carried out for an arbitrary but fixed value of $\lambda$ about the $\tau=0$ limit. The phase transition from phase 1 to phase 2 can be induced by increasing $\lambda$ or $\tau$.
} 
\label{fig:phase_diagram_tau}
\end{figure}
In contrast to the purely perturbative LCE approaches, NLCEs yield numerical data sequences for a fixed value of $\lambda$. Since each cluster is treated non-perturbatively, no perturbative \mbox{parameter} limiting the applicability exists. Yet, only a finite set of clusters can be treated numerically which sets a characteristic length scale $\mathcal{L}$ of quantum fluctuations captured. Consequently, if the physical system has a finite correlation length, the NLCE converges as long as $\xi\sim\mathcal{L}$. Specifically at quantum critical points with $\xi\rightarrow \infty$, one must rely on extrapolation techniques.

One type of extrapolation of NLCE results relies on an appropriate scaling in $\mathcal{L}$ \cite{Kallin2013}. A challenge of these scalings is the assignment of a length scale to a given cluster in more than one dimension. It is possible to extract universal properties relying on scaling arguments. It seems, at least in principle, possible to extract the critical exponent $\nu$ in a similar fashion. The other extrapolation relies on the $\epsilon$-Wynn and related methods \cite{Rigol2007_1,Tang2013}. This method does not require a length scale, yet, this approach does not provide any critical exponents.

In a work of Bernu and Misguich \cite{Bernu2001}, the expansion for entropy in the inverse-temperature variable is converted to an expansion for entropy in the internal energy, allowing to build the ground-state energy and low-temperature power-law behavior into the extrapolation of high-temperature expansion. Our reasoning is in a similar direction.

We suggest a scheme, addressing specifically the power-law behavior close to quantum criticality. The fundamental idea behind the approach is a reformulation of the data sequences as a series expansion in a pseudo parameter, allowing to utilize standard series expansion extrapolation techniques to extract critical properties (see last subsection). The reformulation does not rely on a length scale, yet, the series extrapolation schemes can be physically motivated. Most importantly, in addition to critical points, this scheme provides access to critical exponents.

Let $K_{N,\lambda}$ denote the physical quantity of interest obtained via an NLCE calculation including up to $N$ supersites for the coupling strength $\lambda$. Here a supersite represents the building block of the graph decomposition used in the NLCE, e.g.~a single site, a dimer or more complicated objects consisting of more elementary sites. The values $K_{N,\lambda}$ are, in principle, expected to converge with increasing $N$ as long as $\lambda<\lambda_{\rm c}$. 

Next, we introduce the parameter
\begin{align}
b_{N-1,\lambda}=K_{N,\lambda}-K_{N-1,\lambda},
\end{align}
representing the contributions specific to $N$-site clusters ($K_{0,\lambda}\equiv0$). We then simply rewrite
\begin{align}
K_{N,\lambda}=\sum_{n\geq 0}^{N-1} b_{n,\lambda} = b_{0,\lambda}+b_{1,\lambda}+\dots+b_{N-1,\lambda}.
\end{align}
and define the function
\begin{align}
G_\lambda(\tau)=\sum_{n\geq 0}^{m} b_{n,\lambda} \tau^n = b_{0,\lambda}+b_{1,\lambda} \tau +\dots+b_{m,\lambda} \tau^{m}\label{G_lambda},
\end{align}
with $m=N_\text{max}-1$. The pseudo parameter $\tau$ functions as a substitute for the missing expansion parameter and one aims at \mbox{$\tilde{G}_\lambda(\tau=1)=\lim_{n\rightarrow \infty} G_\lambda(\tau=1)$}. It is therefore possible to apply the extrapolation techniques presented in the last subsection for this high-order series in the parameter $\tau$.

Indeed, the extrapolation techniques applied to data sequences in Ref.~\onlinecite{Rigol2007_1} and Ref.~\onlinecite{Tang2013} are identical to Pad\'e extrapolations evaluated at $\tau=1$. However, here we argue that Dlog-Pad\'e extrapolations are perfectly suited for this purpose if the system is close to quantum criticality.

Following Eq.(\ref{G_lambda}), the results of an NLCE expansion can be interpreted as an expansion in $\tau$ about the local cluster limits $\tau=0$. But in addition to that, the cases $\tau=1$ and $\tau\neq 1$ can be identified with the same physical system in an extended parameter space as visualized in Fig.~\ref{fig:phase_diagram_tau}. In NLCEs, the local cluster limit is closely related to the perturbative limit and the expansion is carried out starting from a specific phase (see for instance the discussion of low-field NLCEs and high-field NLCEs in Ref.~\onlinecite{Kallin2013}). Due to the close relation between the limits $\tau \rightarrow 0$ and $\lambda \rightarrow 0$, it is suggestive that the same phase transition from phase 1 to phase 2 is induced by increasing $\lambda$ or $\tau$ respectively and no intermediate phases occur in $\tau$. If this assumption does not hold, the NLCE approach seems overall problematic.

For values of $\lambda$ close to criticality, one expects that $\tau_{\rm c}$ is close to one, i.e., $G_\lambda(\tau)$ must be evaluated close to criticality and it is therefore reasonable to apply Dlog-Pad\`e extrapolations to obtain approximations of $\tilde{G}_\lambda(\tau=1)$. If the extrapolation yields $\tau_{\rm c}<1$ ($\tau_{\rm c}>1$), one deduces $\lambda>\lambda_{\rm c}$ ($\lambda<\lambda_{\rm c}$). Thus, a scheme of the form $\lambda_{i+1}=\frac{\lambda_i}{\tau_{c,i}}$ allows an iteration to determine $\lambda_{\rm c}$.

Most importantly, due to the universality of the critical exponents, it is possible to extract the critical exponent by applying Eq.~\eqref{extract_exponent} or \eqref{biasnue} with \mbox{$G_\lambda(\tau)\mathrel{\hat=}F(\lambda)$}, at least if $\lambda$ is in the vicinity of $\lambda_{\rm c}$. We stress that this extrapolation scheme does not require any additional numerical overhead and relies solely on the data available. 
%
\section{Application}
\label{Sect:applications}

\begin{figure}
\begin{center}
\includegraphics*[width=0.9\columnwidth]{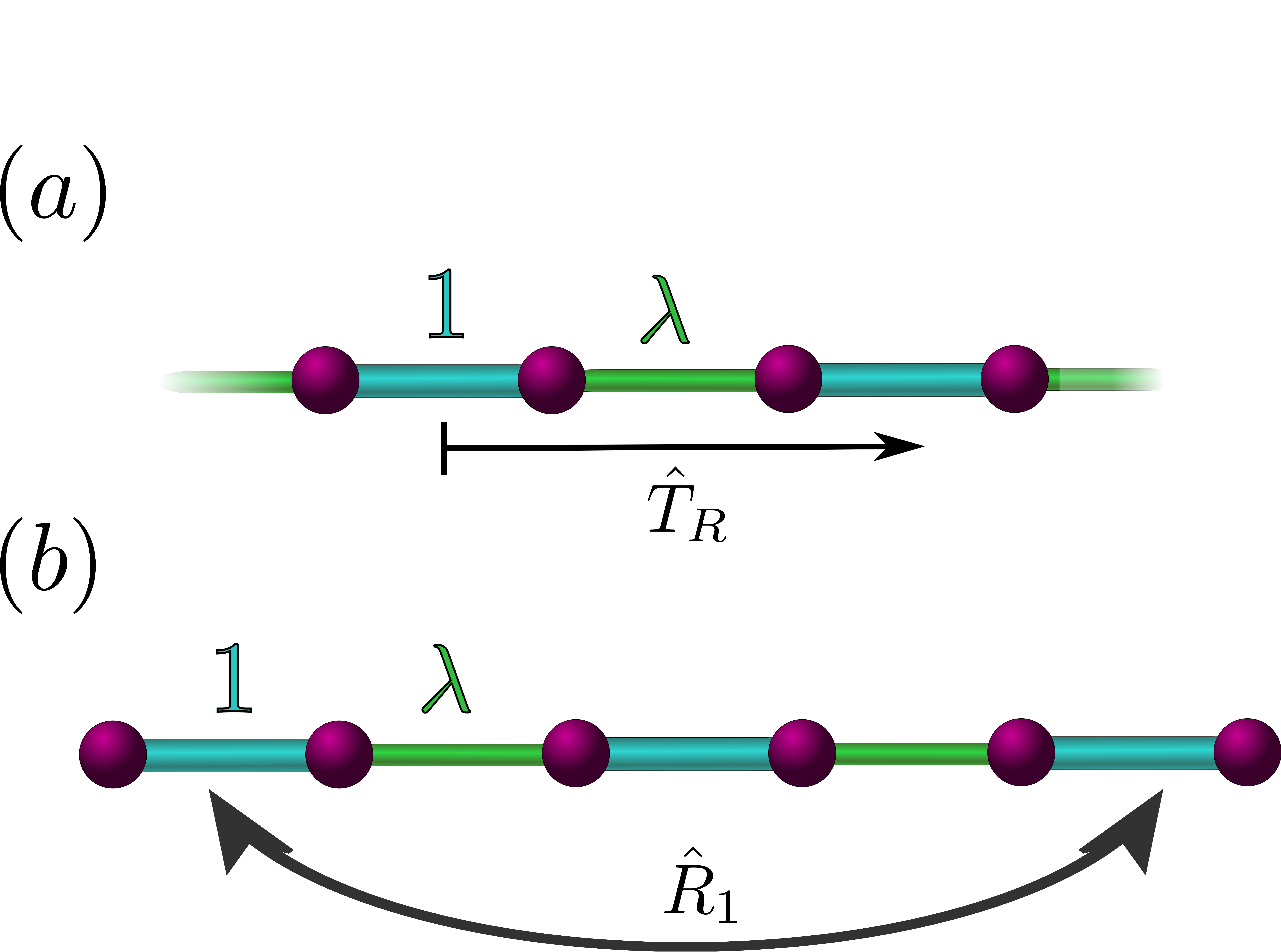}
\end{center}
\caption{(a) Illustration of the (dimerized) Heisenberg chain. The spins 1/2 are skteched as filled circles and the intra-dimer (inter-dimer) couplings by dark (light) lines. The Hamiltonian \eqref{H_dim_chain} is studied about the dimerized limit $\lambda=0$. The system exhibits the translational symmetry $\hat{T}_\text{R}$. (b) Illustration of a chain segment with $N=3$ dimers. The reflection symmetry $\hat{R}_1$ is exploited during the gCUT calculation.
} 
\label{fig:dim_chain_model}
\end{figure}

To demonstrate the applicability of the extrapolation techniques, we study the dimerized Heisenberg spin 1/2 chain, which is depicted in Fig.~\ref{fig:dim_chain_model}(a), using gCUTs. The Hamiltonian is given by
\begin{align}
\mathcal{H}=\sum_{\langle i,j\rangle }{\bf S}_i \,{\bf S}_j+\lambda\sum_{\langle i,j\rangle^\prime}{\bf S}_i \,{\bf S}_j\quad ,\label{H_dim_chain}
\end{align}
where the first (second) sum represents the intra-dimer (inter-dimer) couplings. The different couplings are illustrated in Fig.~\ref{fig:dim_chain_model}(a). The system is gapped for all values $\lambda<1$. Here the ground state is adiabatically connected to the product state of singlets in the limit $\lambda=0$ and the elementary excitations are so-called massive triplons \cite{Schmidt2003} with total spin one which correspond to dressed triplet excitations. At the quantum critical point $\lambda=1$, the model is exactly solvable by Bethe Ansatz \cite{Bethe1931,Hulthen1939,Yang1966_1,Yang1966_1}. Moreover, the corresponding excitation spectrum is exactly known \cite{Cloizeaux1962} and is constituted by gapless fractional spinon excitations carrying a spin 1/2 \cite{Faddeev1981}. This transition is known to be a confinement-deconfinement transition where the triplon gap closes as $(1-\lambda)^{z\nu}$ with the critical exponent $z\nu=2/3$ \cite{Cross1979,Affleck1989}.  

Introducing triplet creation and annihilation operators $\hat{t}^\dagger_{i,\alpha}$ ($\hat{t}^{\phantom{\dagger}}_{i,\alpha}$) with magnetization $\alpha=\{-1,0,1\}$ on dimer $i$, one can rewrite the Hamiltonian as \cite{Knetter2000}
\begin{align}
\mathcal{H}=\tfrac{3}{4}N+\hat{Q}+ \lambda\sum_{n=-2}^{n=2} T_n\quad ,
\end{align}
where $N$ denotes the number of dimers, $\hat{Q}$ counts the number of triplets and $T_m$ ($T_{-m}$) creates (annihilates) in total $m$ triplets on neighboring dimers. The explicit local matrix elements are given in Ref.~\onlinecite{Knetter2000}.

Here we apply gCUTs to derive an effective triplon-conserving model $\mathcal{H}_{\text{eff}}$ with $[\mathcal{H}_{\text{eff}},\hat{Q}]=0$ similarly as in Ref.~\onlinecite{Coester2015} for the triplon excitations of the antiferromagnetic two-leg ladder. The low-energy Hamiltonian can be expressed as
\begin{eqnarray}
 \mathcal{H}_{\text{eff}} &=& \tilde{E_0}(\lambda) + \sum_{i,\delta,\alpha} \tilde{a}_\delta(\lambda)\, \hat{\tilde{t}}^\dagger_{i+\delta,\alpha}\hat{\tilde{t}}^{\phantom{\dagger}}_{i,\alpha} + {\rm h.c.}\nonumber\\
 &+\ldots 
\label{H_eff_set_up_dim_chain}
\end{eqnarray}
where $\hat{\tilde{t}}^\dagger_{i,\alpha}$ ($\hat{\tilde{t}}^{\phantom{\dagger}}_{i,\alpha}$) creates (annihilates) a triplon with magnetization $\alpha$ on dimer $i$. The $\ldots$ refer to other quasi-particle conserving operators corresponding to interactions between triplons which we do not consider in this work. Due to the $SU(2)$ invariance, the hopping elements $a_\delta(\lambda)$ are independent of the triplon flavor $\alpha$. A Fourier transformation then yields the one-triplon dispersion $\omega(k)$. Here, we are specifically interested in the one-triplon gap \mbox{$\Delta\equiv\omega(k=\pi)$} at the quantum critical point $\lambda=1$.

The gCUT is carried out using chain segments of dimers as depicted exemplarily in Fig.~\ref{fig:dim_chain_model}(b) for $N=3$. Using a basis truncation similar to Ref.~\onlinecite{Ixert2014} and exploiting the reflection symmetry $\hat{R}_1$ on each cluster, it is possible to reach graph sizes up to $N=12$ dimers. One therefore restricts the maximal number of basis states $d_{\rm max}$ on graphs to a specific value and compares the results when varying  $d_{\rm max}$. The gCUT is performed at the critical point $\lambda=1$, making a proper description of the system very challenging. On a single cluster, a lot of states merge with the low-energy spectrum and the distinction between genuine and defective interactions, as demonstrated in Ref.~\onlinecite{Coester2015}, becomes difficult. This can be also seen in our calculation where we have chosen the two values $d_{\rm max}=150$ and $d_{\rm max}=200$ as the maximal number of basis states on graphs. The little differences between both calculations can be attributed to this difficulty. 
\begin{figure}[ht!]
\begin{center}
\includegraphics*[width=0.9\columnwidth]{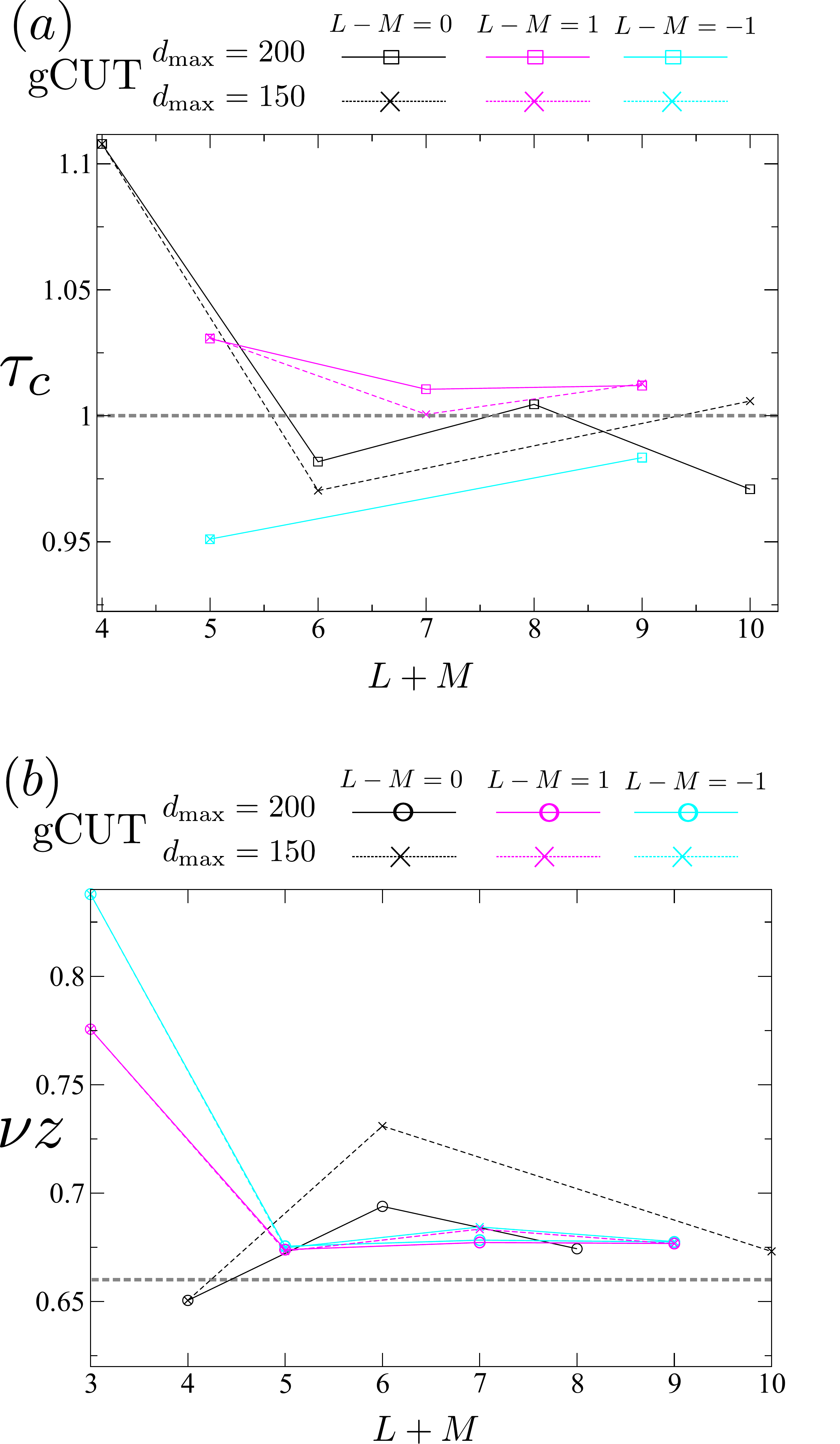}
\end{center}
\caption{(a) Critical points $\tau_{\rm c}$ of $\Delta_{\lambda=1}(\tau)$ obtained via Dlog-Pad\`e extrapolations of the gCUT results. (b) Critical exponents $\nu z$ of $\Delta_{\lambda=1}(\tau)$ obtained via biased Dlog-Pad\`e extrapolations of the gCUT results. The gCUT calculations has been done for $d_{\rm max}=150$ and $d_{\rm max}=20$. The dashed grey lines represent the analytical result.} 
\label{fig:dimerized_chain_critical_values}
\end{figure}

To perform an extrapolation for the numerical data sequence up to $N=12$ of the one-triplon gap at $\lambda=1$, we reformulate this data sequence as an expansion in $\tau$ yielding an order $m=11$ polynomial of the form Eq.~\eqref{G_lambda}. The aim is to extrapolate the resulting series $\Delta_\lambda(\tau)$ to infinite order. We argue that the series should vanish like $\Delta_{\lambda}\propto(\tau_{\rm c}-\tau)^{\nu z}$, making Dlog-Pad\`e extrapolations the method of choice.

The resulting critical values $\tau_{\rm c}$ obtained via Dlog-Pad\`e extrapolations of $\Delta_\lambda(\tau)$ ($\lambda=1$) are shown in Fig.~\ref{fig:dimerized_chain_critical_values}(a). Defective extrapolants are omitted and we consider only the three families with $|L-M|\leq 1$. Since the calculation is performed at the critical point, one expects $\tau_{\rm c}\approx 1$. Indeed, this is consistent with the convergence of the obtained critical values.

Next, we consider the extraction of the critical exponent $\nu z$. We therefore use biased Dlog-Pad\`e extrapolation and implement Eq.~\eqref{biasnue} with $\tau_{\rm c}=1$. The resulting critical exponents are depicted in Fig.~\ref{fig:dimerized_chain_critical_values}(b). Interestingly, our results are in agreement with the known critical exponent $\nu z=2/3$ \cite{Cross1979,Affleck1989} within two percent, which is for a sophisticated quantity like a critical exponent fairly accurate.

\section{Conclusions}
\label{Sect:Conclusions}
We introduced a novel scheme to extrapolate the numerical data sequences which are typically obtained by NLCEs and related numerical techniques. While the mapping of the data sequences to a series in a pseudo parameter changes the perspective, our approach is specifically designed to describe systems close to or even at criticality. Most importantly, this scheme allows to extract critical exponents.

A transfer to other NLCE schemes, for instance to determine the entanglement entropy \cite{Kallin2013}, is straight forward.  The determination of critical exponents would be specifically interesting for the determination of possibly new universality classes for systems not described by Landau's theory similar to Ref.~\onlinecite{Schulz2013}. Additionally, it is appealing to extend these \mbox{considerations} to time-dependent properties calculated via NLCEs \cite{Rigol2014_1} or many-body (de)localization \cite{Devakul2015}.

Moreover, the same line of reasoning can be applied to other numerical techniques yielding also sequences of numerical data points with a similar relation between the limits $\tau \rightarrow 0$ and $\lambda \rightarrow 0$. One example are non-perturbative real-space CUTs in operator space \cite{Krull2012}. Furthermore, it would be interesting to investigate if the scheme can be also applied successfully to variational tensor network calculations.

\end{document}